\documentclass[12pt]{article}
\usepackage[dvips]{graphicx}
\begin{document}

\title{Corrections to a mean number of droplets in
nucleation}

\author{Victor Kurasov}

\maketitle

\begin{abstract}

Corrections to a mean number of
droplets appeared in the process of
nucleation have been analyzed. The
two stage model with a fixed boundary
can not lead to a write result. The
multi stage
generalization of this model also can
not give essential changes  to the two
stage model. The role of several first
droplets have been investigated and it
is shown that an account of only first
droplet with further appearance in
frame of the theory based on the
averaged characteristics can lead to a
suitable results. Both decay of
metastable phase and smooth variations
of external conditions have been
investigated.
\end{abstract}

\section{Introduction}

Up to the last years
a kinetic descriptions  of nucleation
processes were based on the averaged
intensity of droplets appearance,
i.e. on the rate of nucleation. Namely the rate of
nucleation is considered as a central
characteristic
of the first order phase transition.
But the supercritical droplets appear in the system
occasionally and this
feature has to be taken into account.
Since \cite{Vest}, \cite{Koll}, \cite{Kolldyn}
several attempts to include
stochastic effects of droplets
appearance were made. Unfortunately,
they could not give an adequate  and rather
precise description of stochastic effects.

The arguments  in
justification of kinetics based on
the averaged characteristics  (see \cite{PhysRevE94})
remain valid even after
stochastic  formulation of
the nucleation problem has been presented.
When the number of droplets in the system is
very big the result of
the theory based on the averaged characteristics
(TAC) is precise. Namely the
number of droplets is the
central
characteristic of the nucleation process and it is
calculated in experiments.
When the number of droplets
appeared in a system isn't so
great (in the free molecular
regime of growth it occurs
only due to a volume
of a system) one can
speak about corrections to a mean
value of the total number
of droplets appeared in the
system in comparison with result predicted by TAC.
This value will be the  main
object of investigation here.

In diffusion regime of the droplets
growth the kinetic
description is based on special models (see
\cite{PhysicaA}) and
there is no need to analyze this
regime here. So, in this paper
only the free molecular
regime of droplets growth will be considered.

In investigation of stochastic effects of nucleation
 one can see rather easy that
the first correction terms
are equal to zero. So, there
appear a problem to calculate the higher terms of
decomposition. It is rather difficult to
perform precise
calculations and we shall be interested at
least to get
estimates for these coefficients
to prove the smallness of
the total amount of corrections. But
even the
calculation
of the very first terms meets  technical
difficulties
(see \cite{Kolldyn}, \cite{Koll}).
One has also to stress that the zero shift
found in these papers was a natural  result
of linearization
made to overcome technical difficulties.
Then it can not be considered there as a
true physical result,
but only the consequence of lineariazation.
So, it is necessary to
propose a  method to calculate
the estimates for
coefficients in asymptotic expansions
due to stochastic
corrections of nucleation.

There are two characteristic situations of external
conditions in which kinetics of nucleation
ordinary was constructed.
These conditions are:
\begin{itemize}
\item
decay of metatsable phase when at
some moment the metastable phase is created
and later there
is no external influence on the system;
\item
smooth variation of
external influence on the system.
\end{itemize}
In both situations
corrections will be established.

The structure of the further analysis
is following:
\begin{itemize}
\item
At first  we shall analyze the two
stage model with a fixed boundary. The
result will be disappointing - one can
not reproduce the results of numerical
simulation. This
corresponds to the difficulties
of this model in prediction of the
value of dispersion.
\item
To improve results we shall use the
multi stage generalization of the last
model. But corrections to the two stage
model will be small and this can not
lead to suitable results.
\item
Then a new approach will be used. We
shall analyze the role of stochastic
appearance of the first droplets.
Results will be very fruitful and one
can see that already account of one
droplet will lead to success.
\item
All these considerations will be made
both for decay of metastable phase and
for the smooth variations of external
conditions.
\end{itemize}

\section{Decay of metastable phase}

The kinetics of nucleation in frames of
the theory based on averaged characteristics (TAC)
 can be described
by the following equation
$$
g(z) =  \int_0^{z} (z-x)^3 \exp(-g(x)) dx
$$
where unknown function $g$ is the
renormalized value of
the number of molecules in a liquid
(new) phase. This
result can be found in \cite{Monodec}.
The meaning of variables $z,x$ can be
also found in
\cite{Monodec}.
Since \cite{Monodec} it is known that one can
describe kinetics in frames of
monosdisperce
apporoximation, i.e.
$$
g(z) = N_{eff}(z) z^3
$$
where
$N_{eff}$ is the effective number of
droplets in monodisperce
peak, namely
$$
N_{eff} (z) = z/4
$$

The monodisperce approximation
can  be chosen as the fixed  (not floating)
monodisperce
approximation (see \cite{Monodec}) and leads
to the following expession for the
size spectrum
$$
f(x)  = f_* \exp(-N_{eff} x^3)
$$
Here $f_*$ is the amplitude of spectrum,
$$
N_{eff} = N(\Delta x / 4)
$$
and
$\Delta x$ is a width of a whole
spectrum (connected with the
duration of a nucleation period).

The total number of droplets
can be obtained on the base of $f$ as
$$
N_{tot} = \int_0^{\infty} dx f(x)
$$

For $N_{eff}$ we have a Gaussian distribution
with standard dispersion since formation of the
first
$N_{eff}$ droplets can be treated as the sequence
of independent events
$$
P(N_{eff} )
\sim
\exp(- \frac{(N_{eff} - <N_{eff}> )^2 }{ 2 <N_{eff}> })
$$
Here $<N_{eff}>$ is the mean value of $N_{eff}$

Then for the averaged
value of $N_{tot}$, i.e. for $<N_{tot}>$ we have
the following formula
$$
<N_{tot}> =
 \int_{-\infty}^{\infty} d N_{eff} P(N_{eff})
f_* \int_0^{\infty} \exp(-N_{eff} x^3) dx
$$

Now with the help of formula
$$
\int dy \exp(-ya) \exp(-c (y-b)^2 ) \sim
\exp(- ba + \frac{a^2}{4 c})
$$
we fulfill integration over $N_{eff}$.
Here $b = <N_{eff}>$, $a=x^3$, $c^{-1} = 2
 <N_{eff}>$.
As the result we have
\begin{equation} \label{tt}
<N_{tot}>  \sim \int_0^{\infty} \exp(-<N_{eff}> x^3 +
\frac{x^6}{2} <N_{eff}> )
dx
\end{equation}
The second term in exponent, i.e. $\frac{x^6}{2}
<N_{eff}>$ is the correction term
which can be seen
from
$$
<N_{tot}>  \sim f_*^{3/4} \int_0^{\infty}
\exp(- y^3 +
\frac{y^6}{2 <N_{eff}>} )
dy
$$

As a rough estimate we can take $y$ in correction
term as $y \approx 1$ and get
$$
<N_{tot}> = <N_{tot\ 0}> \exp(\frac{1}{2 <N_{eff}> })
$$
where $<N_{tot\ 0}>$ is the value $N_{tot}$
calculated without
stochastic effects taken into account,
i.e. in frames of
TAC. Then for this value one can get expression
$$
<N_{tot\ 0}>  \sim \int_0^{\infty}
\exp(-<N_{eff}> x^3  )
dx
$$
Having  noticed that
$$
<N_{eff}> \approx <N_{tot\
0 } >/ 4
$$
we get
$$
<N_{tot}> = <N_{tot\ 0}> \exp(\frac{2}{ <N_{eff}>
})
$$
Decomposition of exponent gives
$$
<N_{tot}> = <N_{tot\  0}> + 2 +
\frac{2}{<N_{tot\ 0}> }
+ ...
$$

Another more balanced variant of
consideration is to use
decomposition of
$$
 \exp(
\frac{x^6}{2} <N_{eff}> ) =
1 +\frac{x^6}{2} <N_{eff}> + \frac{x^{12}}{8}
<N_{eff}>^2
$$
already in (\ref{tt}).
At least the integral then will have no problems
with convergence.
We have
$$
<N_{tot}>  \sim \int_0^{\infty} \exp(-<N_{eff}> x^3
)(
1 +\frac{x^6}{2} <N_{eff}> + \frac{x^{12}}{8}
<N_{eff}>^2
)
dx
$$
Integration can be fulfilled separately for every
term in decomposition.
Then we come to
$$
<N_{tot}> = <N_{tot\ 0}> + A_1 + \frac{A_2}
{<N_{tot\ 0}>}
+ ...
$$
where constants $A_1$ and $A_2$ are given by
% fail addition1.mws
$$
A_1 = 2 \int_0^{\infty} \exp(-y^3) y^6 dy /
\int_0^{\infty}
\exp(-y^3)  dy
= 8/9
$$
$$
A_2 = 2 \int_0^{\infty} \exp(-y^3) y^{12} dy /
 \int_0^{\infty}
\exp(-y^3)  dy
= 510/89 = 6.91
$$

From the functional forms of expressions for
$A_i$ one can
see that $A_1$ is determined rather smart
while the error
in $A_2$ can be essential. The reason is
the rapidly growing
term $y^{12}$ in subintegral function.
Already $y^{6}$ in
expression for $A_1$ grows too rapidly.
So, in subintegral
functions the main role belong to $y$
corresponding to
droplets appeared at the very end
of the nucleation period.
But the form of spectrum $\sim \exp(-y^3) $
is determined at
the
back side of spectrum (i.e. at $y > 1$ )
with a low accuracy.
 In
TAC the weight of such droplets was
negligible and the result
was accurate.
Here the error can be essential. That's why it
is reasonable to restrict
the decomposition only by
the first term $A_1$.

In Figure 1 one can see the results
of numerical simulation
(oscillating curve) and
analytical approximation (smooth
monotonuous
curve) for the relative value
$$
P = <N_{tot}> / <N_{tot\ 0}> -1
$$

% fail addition2a.mwz

\begin{figure}[hgh]

\includegraphics[angle=270,totalheight=10cm]{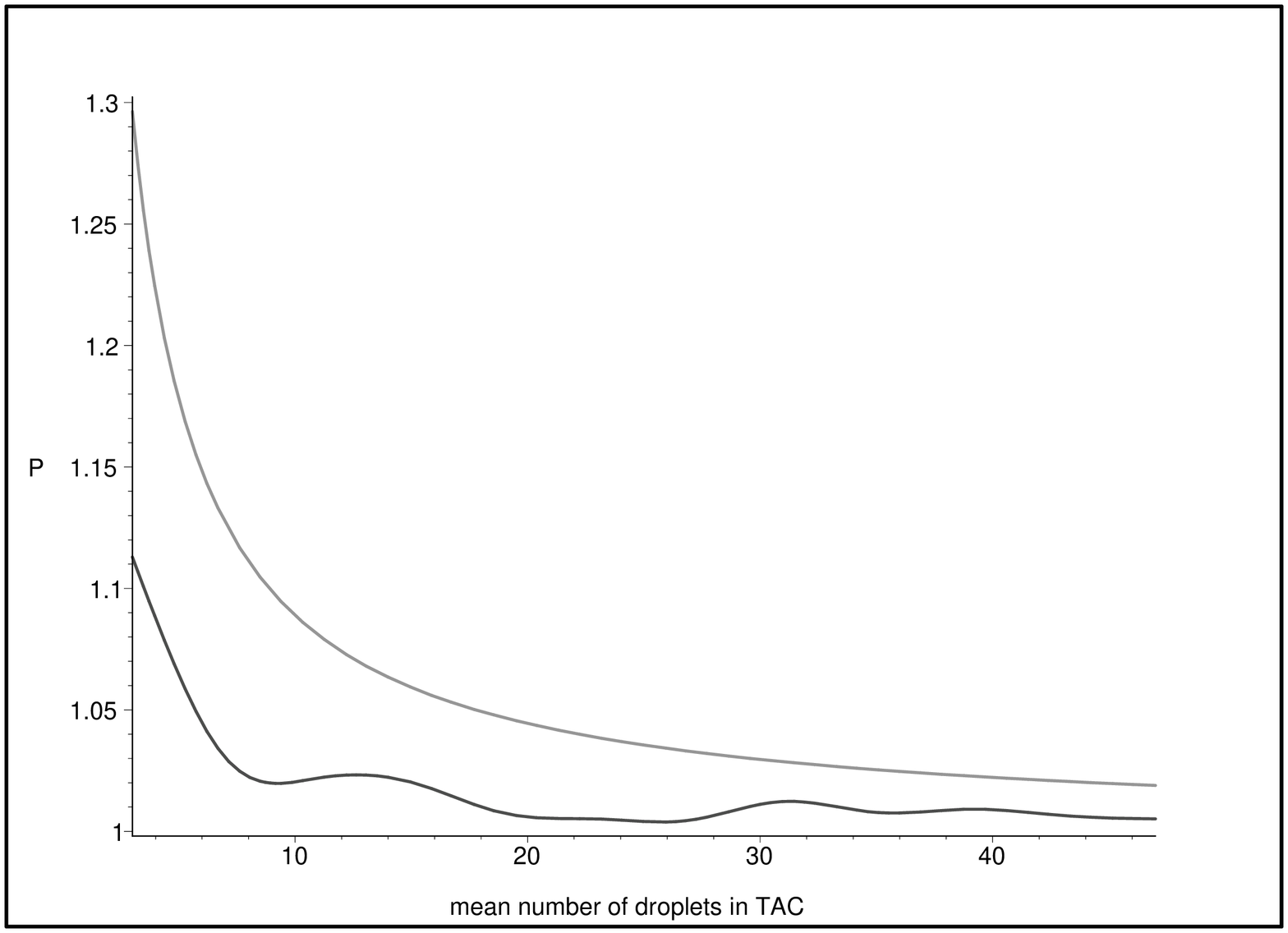}

\begin{caption}
{
Numerical and analytical solutions in the situation of
decay. The initial monodisperce approximation is
considered.
}
\end{caption}
\end{figure}

One can see that there is no satisfactory coincidence
between the theoretical result and the result of
simulation. The reason is the roughness
of monodisperce
approximation used in \cite{Monodec} and applied
here.

Now we shall take a more
refined approximation used to
calculate the value of
dispersion initiated by stochastic
appearance of droplets in the process of decay
\cite{Decaydispersion}. This approximation is the
following: the
length of formation of monodisperce spectrum
is $2*l$ where $l=0.2$; the  monodisperce
spectrum is formed at $2*l - b$ where $b=0.336$. The
derivation of this approximation can be found in
\cite{Decaydispersion}.
Here
$$
<N_{tot\ 0}> = 2l -b + \int_0^{\infty}
\exp(-2lx^3) dx
$$

Then one has to reconsider the value of $A_1$.
It can be done only rather
approximately.
One
can  see that $A_1$ is  reciprocal to
the total number of droplets in
monodisperce
spectrum which is now $2*l = 0.4$ instead of
$0.25$ in initial monodisperce approximation.
Then instead of previous
$A_1$ one has to take
$$
A_1 \rightarrow
 A_1  0.25 /0.4 = 1/3.6
$$
The results are shown in Figure 2

% fail addition2b.mwz

\begin{figure}[hgh]

\includegraphics[angle=270,totalheight=10cm]{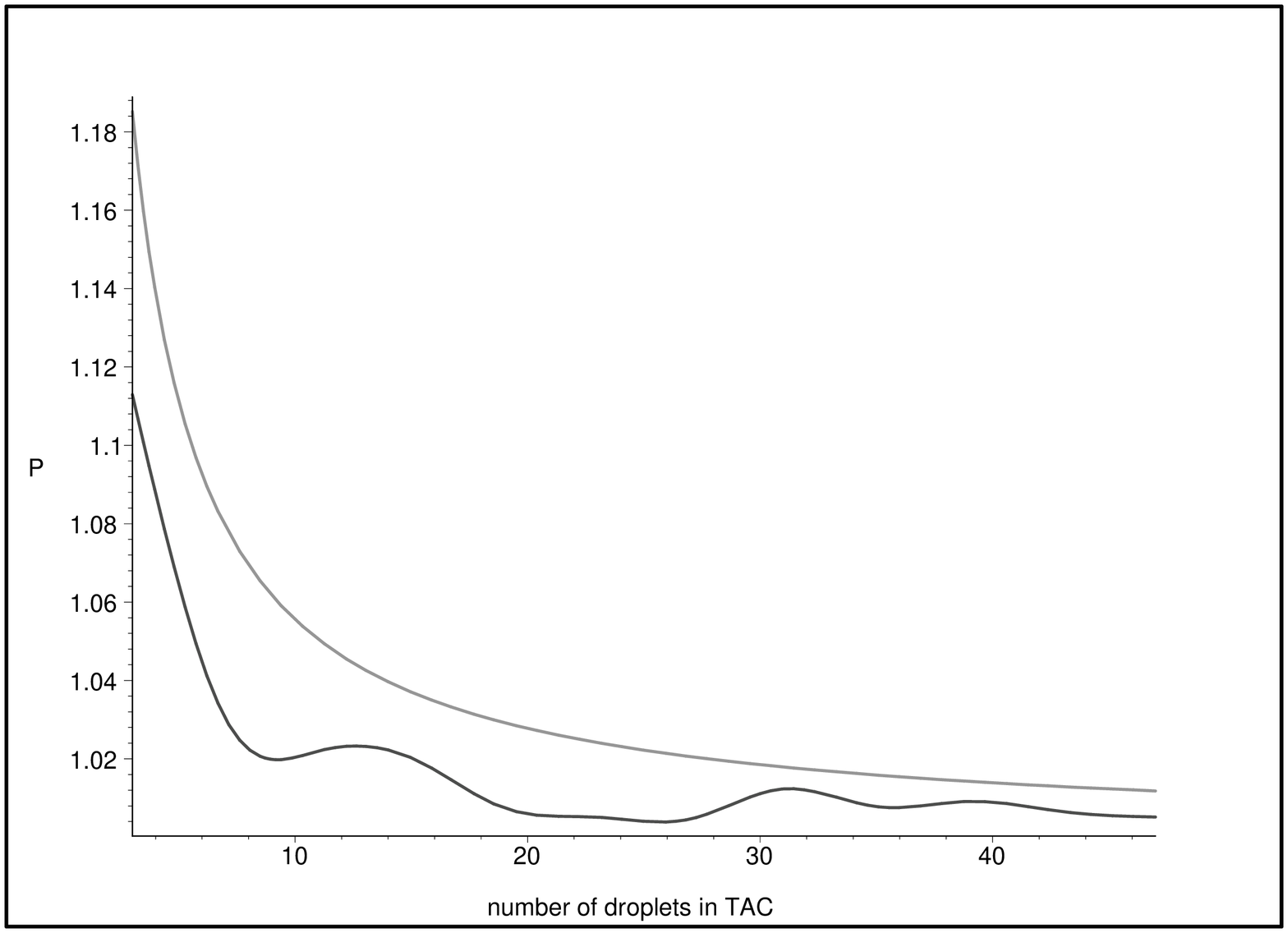}

\begin{caption}
{
Numerical and analytical solutions
in the situation of
decay. The shifted monodisperce
approximation is
considered.
}
\end{caption}
\end{figure}

Here the coincidence between curves became better
but it is not still  satisfactory. This
means that the model with a fixed
boundary can not give a good result.
This corresponds to the difficulties in
calculation of dispersion on the base
of the model with a fixed boundary
\cite{statiae}.

\section{Smooth variations of external conditions}

In the situation of smooth variation of external
conditions (so called dynamic conditions) we can
fulfill the same procedure.

The evolution equation in TAC looks like
$$
g = c^{-1} \int_{-\infty}^{\infty} (z-x)^3 \psi(x)
\exp(-g(x)) dx
$$
where $\psi = \exp(x)$ describes the
change of external conditions
and the renormalization to cancel the coefficient
in $\psi$ is used. As a compensation
for such renormalization
the coefficient
$c=0.189$ or $c=1/6$
(it depends on the type of choice of
the base of decompositions (see \cite{PhysRevE94}))
appears.

Here we shall use the
monodisperce approximation also.
The monodisperce approximation was proposed in
\cite{Monodyn} where all details can be found.
In the fixed monodisperce approximation one can
write
$$
f = f_* \exp(x- <N_{eff}> (x+3)^3)
$$
for the spectrum of droplets sizes.
Here the effective number of droplets is given
by
$$
N_{eff} = \frac{6 c}{27} , \ \ \ c=0.189
$$
or
$$
N_{eff} = \frac{1}{27}
$$

For the total number of droplets we have an
evident expression
$$
N_{tot} = f_* \int_{-\infty}^{\infty} dx
\exp(x- <N_{eff}> (x+3)^3 )
$$

For the mean total number of droplets one can  get
$$
<N_{tot}> = f_* \int d N_{eff} \int_{-\infty}^{\infty} dx
\exp(x- <N_{eff}> (x+3)^3 )
P(N_{eff})
$$
where $P(N_{eff})$ is the distribution function
for the quantity of effective droplets.

For $P(N_{eff})$ we have an evident Gaussian
distribution  with dispersion of ideal gas
$$
P(N_{eff}) \sim \exp( - \frac{(N_{eff} -
<N_{eff}>)^2}{2 <N_{eff}>}
$$

Having fulfilled integration one gets
$$
<N_{tot}> \sim f_* \int_{-\infty}^{\infty}
dx \exp(x - <N_{eff}> (x+3)^3)
\exp(\frac{(x+3)^6}{2} <N_{eff}>)
$$

Certainly, the last integral doesn't
converge. We need a regularization
which will be done below.

Then one has to decompose
$$
\exp(\frac{(x+3)^6}{2} <N_{eff}>)
= 1 +\frac{(x+3)^6}{2} <N_{eff}>
+ \frac{(x+3)^{12}}{8} <N_{eff}>^2 + ...
$$
and then one can fulfill integration for every
term. Now the integral has no problems with
convergence.

In above formulas $<N_{eff}> = 1/27$
which is a natural
requirement to use the
monodisperce approximation. To
calculate correction terms one has to
 include
 explicitly the volume of the
 system $V$ (i.e. the real mean
number of droplets).
Now we shall give the
explicit formulas for correction terms.
We have to calculate the  value
$$
<N_{tot}> =
\int_{-\infty}^{\infty} dx \int_{-\infty}^{\infty}
\exp(x-\frac{N_{eff}}{V} (x+3)^3 )
\exp(-\frac{(N_{eff} -
\tilde{N_{eff}})^2}{2 \tilde{N_{eff}}} d
N_{eff}
$$
$$
\tilde N_{eff} \equiv < N_{eff} >
$$
Earlier we decomposed
$\exp(-\frac{(N_{eff} - \tilde{N_{eff}})^2}{2
\tilde{N_{eff}}})$
and had some problems with convergence. Now we
shall decompose
$\exp(x-\frac{N_{eff}}{V} (x+3)^3 )$.
At first we shall present
this exponent as
$$
\exp(x-\frac{N_{eff}}{V} (x+3)^3 ) =
\exp(x-\frac{\tilde{N_{eff}}}{V} (x+3)^3 )
\exp(-(\frac{N_{eff}}{V} -
\frac{\tilde{N_{eff}}}{V}) (x+3)^3 )
$$
The decomposition of the last exponent gives
$$
\exp(-(\frac{N_{eff}}{V} -
\frac{\tilde{N_{eff}}}{V}) (x+3)^3 ) =
1+ (x+3)^3 \frac{N_{eff} - \tilde{N_{eff}}}{V} +
\frac{(x+3)^6}{2}\frac{(N_{eff} -
\tilde{N_{eff}})^2}{V^2}+
$$
$$
\frac{(x+3)^9}{6}\frac{(N_{eff} -
\tilde{N_{eff}})^3}{V^3}+
\frac{(x+3)^{12}}{24}
\frac{(N_{eff} - \tilde{N_{eff}})^4}{V^4}
$$

The calculation of integrals gives
$$
\int_{-\infty}^{\infty} \exp(-\frac{(N_{eff} -
\tilde{N_{eff}})^2}{2 \tilde{N_{eff}}} ) d N_{eff} =
\sqrt{\pi} (2 \tilde{N_{eff}})^{1/2}
$$
$$
\int_{-\infty}^{\infty} \exp(-\frac{(N_{eff} -
\tilde{N_{eff}})^2}{2 \tilde{N_{eff}}} )
(N_{eff} - \tilde{N_{eff}})^2 d N_{eff} =
\frac{1}{2} \sqrt{\pi} (2 \tilde{N_{eff}})^{3/2}
$$
$$
\int_{-\infty}^{\infty} \exp(-\frac{(N_{eff} -
\tilde{N_{eff}})^2}{2 \tilde{N_{eff}}} )
(N_{eff} - \tilde{N_{eff}})^4 d N_{eff} =
\frac{3}{4} \sqrt{\pi} (2 \tilde{N_{eff}})^{5/2}
$$

We have to notice that $\tilde{N_{eff}} = V/27$.
Then we have the decomposition
$$
<N_{tot}> = <N_{tot} (V=\infty) >
(1 + \frac{1}{2 V^2}  \tilde{N_{eff}}
\frac{
\int_{-\infty}^{\infty}
\exp(x - \frac{1}{27} (x+3)^3) (x+3)^6 dx
}
{
\int_{-\infty}^{\infty}
\exp(x - \frac{1}{27} (x+3)^3)  dx
}
$$
$$
+
\frac{1}{8 V^4}  \tilde{N_{eff}}^2
\frac{
\int_{-\infty}^{\infty}
\exp(x - \frac{1}{27} (x+3)^3) (x+3)^{12} dx
}
{
\int_{-\infty}^{\infty}
\exp(x - \frac{1}{27} (x+3)^3)  dx
})
$$

Now we have to note that the lower limit of
integrations has to
be put $x=-3$ because the monodisperce
approximation begins to
work only at $x>-3$.
The region $x <3$ has
negligible influence in the total
amount of droplets.
Then
$$
<N_{tot}> = <N_{tot} (V=\infty) >
(1 + \frac{1}{2 V^2} \frac{2 \tilde{N_{eff}} }{2}
\frac{
\int_{-3}^{\infty}
\exp(x - \frac{1}{27} (x+3)^3) (x+3)^6 dx
}
{
\int_{-3}^{\infty}
\exp(x - \frac{1}{27} (x+3)^3)  dx
}
$$
$$
+
\frac{3}{24 V^4}  \tilde{N_{eff}}^2
\frac{
\int_{-3}^{\infty}
\exp(x - \frac{1}{27} (x+3)^3) (x+3)^{12} dx
}
{
\int_{-3}^{\infty}
\exp(x - \frac{1}{27} (x+3)^3)  dx
}
$$

Here there were no problems with
convergence.
We can calculate
the integrals numerically which gives
% grupppa program "my"
$$
<N_{tot}> =
<N_{tot} (V= \infty) >
(1+ \frac{A_1}{<N_{tot} (V= \infty) >} +
\frac{A_2}{<N_{tot} (V= \infty) >^2} + ...
$$
$$
A_1 =
\frac{1}{2*27^2} \frac{\int_{-\infty}^{\infty}
\exp(x-\frac{1}{17} (x+3)^3 ) (x+3)^6 dx}
{\int_{-\infty}^{\infty}
\exp(x-\frac{1}{17} (x+3)^3 )  dx}
= 1.7
$$
$$
A_2 =
\frac{1}{8*27^4} \frac{\int_{-\infty}^{\infty}
\exp(x-\frac{1}{17} (x+3)^3 ) (x+3)^{12} dx}
{\int_{-\infty}^{\infty}
\exp(x-\frac{1}{17} (x+3)^3 )  dx}
= 8.1
$$
This is the final result.

Again one has
to note that $A_2$ is determined with
uncertainty caused by approximate
description of the back
side of spectrum. Here $A_2$ is
calculated to see that
there is no singularities in decomposition.
Again it is
reasonable to take into account only the first
correction term.

In the situation of the
smooth behaviour of external
conditions there is no moment of start.
The point of formation of monodisperce
spectrum is $z=-3$ and it is
determined in the internal point.
So, when we observe the
subintegral
function it has a maximum not in the boundary
point (as
in decay when it is $z=0$) but in an internal
point near $z=-3$. The subintegral function
$(z-x)^3 \psi(x)
\exp(-g(x))$
is rather symmetric around $x=-3$ (in decay
one can not imagine that  the subintegral function
is
symmetric around the boundary point).

Numerical simulation  and analytical result
  can be seen
in Figure 3.
% fail addition3.mwz

\begin{figure}[hgh]

\includegraphics[angle=270,totalheight=10cm]{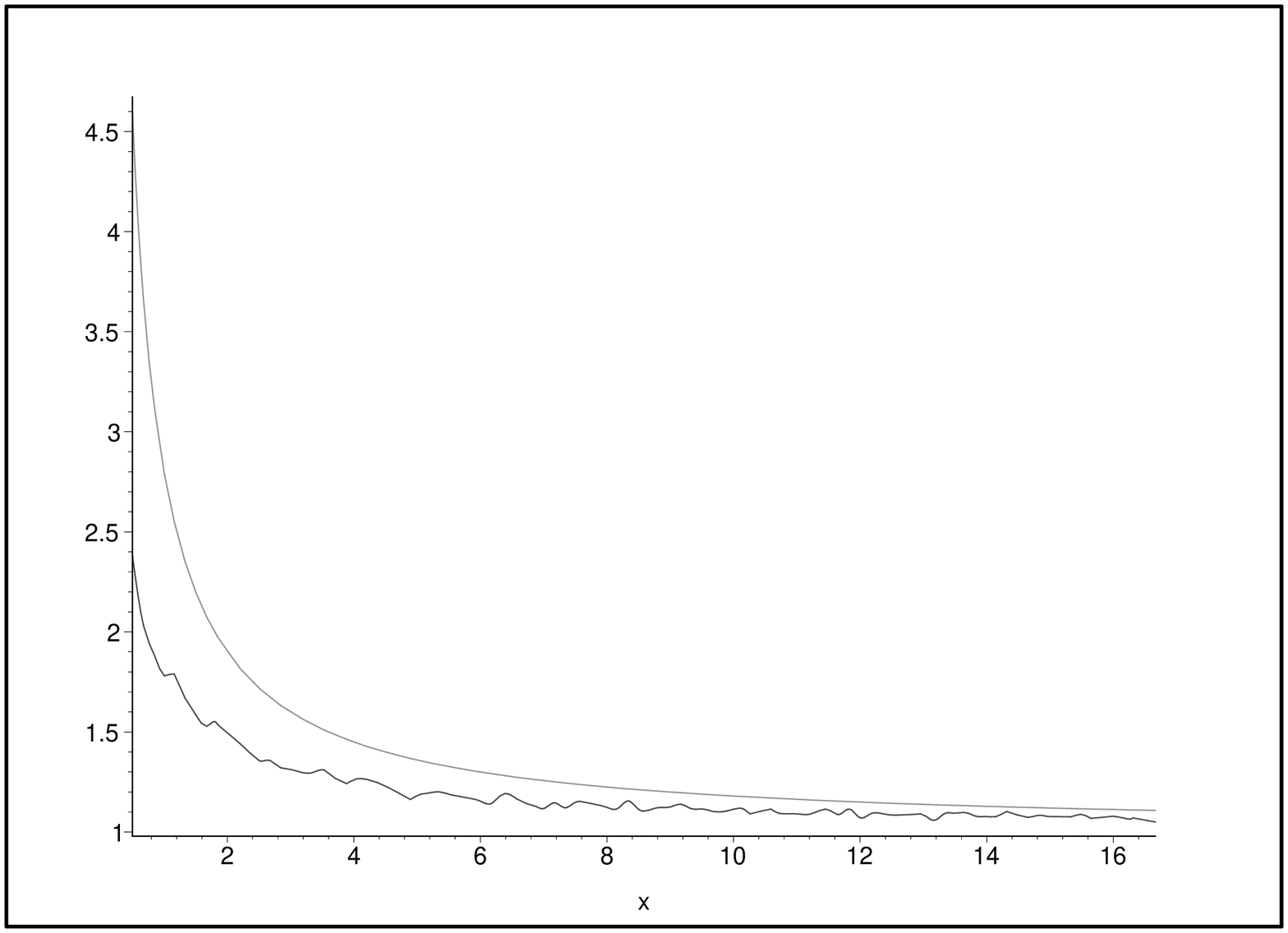}

\begin{caption}
{
Numerical simulation and analytical
solution under the
smooth external conditions. The
dependence of $P$ over the
volume of the system $V \equiv x $ is plotted.
}
\end{caption}
\end{figure}

One can  not see the satisfactory coincidence
between
theoretical result (monotonuous curve)
and numerical
siulation (oscillating curve).
So, the  used model can not give the
good results.

To the data of numerical simulation
one can suggest a phenomenological approximation
$$
N \sim V (1 +\frac{A}{V}+ ... )
$$
$$
A=5/6
$$

One can establish the universality also
in stochastic
formulation of the problem. Then the parameter $A$
is no more than a universal constant.
The numerical simulation
 gives $A=5/6$. Generally speaking one
 can stop here
all investigations.

So, the model  with a fixed boundary
failed in determination of corrections
to the number of droplets. One has to
turn to the models with floating
boundary which was succesfully applied
in \cite{Decaydispersion}, \cite{floatingdindisp}
to the calculation of dispersion. But
in the models with a floating boundary
the correction to the mean number of
droplets is zero. This isn't the error
of the model but the level of
description is limited here.

The next step in consideration is to
analyze whether one can reconsider the
previous results . The fact is that in
nucleation one can see some balancing
forces. This can require the addition
more detailed analysis containing
several stages model instead of two
stages.

\section {Three stage scheme}

In nucleation under the smooth behavior of external
conditions one can observe the specific property of
compensation  which can damage the
previous consideration.
One can analytically
observe the fact of compensation
briefly described below
which  means that one can not use two
stage scheme to get
corrections for the mean number of droplets.
The result of the three cycle scheme can be the
leading term
in the shift of the mean droplet number.
So, we have to use
the three stage scheme.

The effect of compensation in the two
stage scheme require to give
estimates in a three stage scheme.
At first we shall explain the
cancellation of effects in the two stage scheme.

The number of
droplets $N_{tot}$ can be in TAC  in the
monodisperce approximation  calculated as
$$
N_{tot}^{mean} =
\int_{-3}^{\infty} \exp(x - N_{eff}(x+3)^3) dx
$$
Here $N_{eff}$ is $1/27$.

If due to stochastic effects the
necessary number $N_{eff}$
appeared  up to the "moment" $z =
-3-\delta$ instead of $z = -3$ then
we have to  calculate
$N_{tot}$ as
$$
N_{tot} (\delta)=
\int_{-3-\delta}^{\infty}
\exp(x - N_{eff}(x+3+\delta)^3) dx
$$
The calculation gives
$$
N_{tot} (\delta)=
\int_{-3}^{\infty}
\exp(-\delta) \exp(y - N_{eff}(y+3+\delta)^3) dx
$$
for
$y=x+\delta$
and
$$
N_{tot} (\delta)= <N_{tot}> \exp(-\delta)
$$

Now we shall establish the distribution
$P(\delta)$ of the shift
$\delta$. The distribution
$P(N_{eff})$ is the ordinary Gausiian
distribution
$$
P(N_{eff}) \sim
\exp(-\frac{(N_{eff} - 1/27)^2}{2/27})
$$
To get $P(\delta)$ we use
$$
P(N_{eff}) d N_{eff} =
P(\delta)  d \delta
$$
The derivative
$ d N_{eff} / d \delta$ is
$$
\frac{ d N_{eff} }{ d \delta} =
\exp(\delta)
$$
Then
$$
P(\delta) = P(N_{eff}) \exp(\delta)
$$
The second factor completely compensates
the shift in the total
number of droplets. Really,
$$
<N_{tot}> =
\int P(\delta) N_{tot}(\delta) d\delta =
N_{tot}^{mean}
$$

This compensation  shows the zero
effect in the shift of droplets number
in the two cycle scheme and  requires to consider
the three stage scheme.

Consider the three stage scheme qualitatively.
We have to mention
that the value $N_{eff}$ doesn't purely appear
under  the ideal
conditions. Already at $z = -3$
the small part of substance is in
the droplets. The main consumers of vapor
at $z = -3 $ are the
droplets appeared at $z = -6$.
Here we can use also the modisperce
approximation and present  $g$ at $z=-3$ as
$$
g(z=-3) =
N_{init} (z +6)^3
$$
with parameter $N_{init}$ of
initial monodisperce approximation.

Since the stochastic number $\hat{N}_{init}$
doesn't coincide with
the value $\bar{N}_{init}$ calculated in TAC
we can see the deviation of stochastic value
$\hat{g}(z=-3)$ from
the value $\bar{g}(z=-3)$
calculated in TAC.

It seems that we come to the
situation which has been already
described in the two stage scheme.
But now we don't observe the
effect of compensation because this effect
takes place only due to
the integration in the infinite limits.
But here such an
integration
is absent - moreover we need the effect up to
the
fixed moment $z=-3$. So, we shall see the effect
which results in
the difference of the mean value $<g(z=-3)>$
from $\bar{g}(z=-3)$.
Then we
see the regular shift $\delta z $ of the moment until
which the
number of droplets $N_0$ appears. This regular shift
leads to the regular
shift in the total number of droplets.

Now we fulfill
the  computations.  The number of
droplets  formed until $z_l = -3$ in TAC is
$$
\bar{N}_{eff} =
\int_{-\infty}^{-3}
\exp(x- k \bar{N}_{init} (x+6)^3) dx
$$
Parameter $k$ has here a role  like $c^{-1}$
in the two cycle scheme had.

The distribution
$P(\hat{N}_{init})$ of the stochastic
number of initial
droplets $\hat{N}_{init}$ has a normal
Gaussian form
$$
P(\hat{N}_{init})
\sim
\exp(- \frac{(\hat{N}_{init} - \bar{N}_{init})^2}{ 2
\bar{N}_{init} }  )
$$
The value of
dispersion here corresponds to the fact that
we have a free stochastic appearing
of droplets.

Here appeared a
special question whether it is possible to
write the gaussian
distribution for the total number of
events (here it is
the number of appeared droplets). But this
question can  solved positively.

Then
$$
<\hat{N}_{eff}  (z=-3) > =
\int
P(\hat{N}_{init})
\exp(x - k \hat{N}_{init} (x+6)^3 ) dx
$$
Hence
$$
<\hat{N}_{eff}  (z=-3) >
\neq
 \bar{N}_{eff}
 $$

Then we can calculate the regular shift
$\delta z$
along $z$-axis for
the moment until  $\bar{N}_{eff}$
droplets appeared.
For the value of $\delta z$ we get
$$
\delta z =
\frac{
\exp(-3) \frac{\bar{N}_{eff} -
<\hat{N}_{eff}>}{\bar{N}_{eff}}
}
{\exp(x|_{x=-3} -
k \bar{N}_{init} (x|_{x=-3} + 6)^3 ) }
$$

Denote by $\delta$ the following value
$$
\delta  =
\hat{N}_{init} - \bar{N}_{init}
$$
Let us calculate $<\hat{N}_{eff}>$
$$
<\hat{N}_{eff}(z=-3)> =
\int_{-\infty}^{\infty}
d \delta
\int_{-\infty}^{-3} dx
\frac{\exp(-{\delta^2}{2 \bar{N}_{init}} V^{-1})}
{\sqrt{2 \pi \bar{N}_{init} V^{-1}}}
\exp(x - k (\bar{N}_{init} +\delta ) (x+6)^3 ) V
$$
Recall that $V$ is
the volume of the system and here we
have to introduce it explicitly.

We fulfill the calculations and get
$$
<\hat{N}_{eff}(z=-3)> =
$$
$$
\int_{-\infty}^{\infty}
d \delta
\int_{-\infty}^{-3} dx
\frac{\exp(-{\delta^2}{2 \bar{N}_{init}} V^{-1})}
{\sqrt{2 \pi \bar{N}_{init} V^{-1}}}
\exp(x - k \bar{N}_{init}   (x+6)^3 ) V
( 1 - k
\delta (x+6)^3 + \frac{k}{2}  \delta^2  (x+6)^6 )
$$

Then the deviation
between $<\hat{N}_{eff}(z=-3)>$ and
$<\bar{N}_{eff}(z=-3)>$ will be
$$
<\hat{N}_{eff}(z=-3)>
 - <\bar{N}_{eff}(z=-3)> =
$$
$$
 \frac{1}{\sqrt{\pi}}
 \int_{-\infty}^{\infty} d y
  \int_{-\infty}^{-3} dx
  \exp(-y^2)
  \exp(x-k \bar{N}_{init} (x+6)^3 )
   \frac{k^2}{2} y^2
   (x+6)^6 (2 \bar{N}_{init} V)^{-1}
$$

Since the monodisperce
approximation becomes suitable only
at $z = -6 $ it is more  reasonable to write
$$
<\hat{N}_{eff}(z=-3)>
 - <\bar{N}_{eff}(z=-3)> =
$$
$$
 \frac{1}{\sqrt{\pi}}
 \int_{-\infty}^{\infty} d y
  \int_{- 6 }^{-3} dx
  \exp(-y^2)
  \exp(x-k \bar{N}_{init} (x+6)^3 )
   \frac{k^2}{2} y^2 (x+6)^6 2 \bar{N}_{init} V^{-1}
$$

Then for the average
number of droplets appeared in the
system
$$
<\hat{N}_{eff}(z=-3)> V
 =  ( \bar{N}_{eff}(z=-3)|_{
\bar{N}_{eff}(z=-3) = \exp(-3)
 }
+ C_0 V^{-1} )
V
$$
where
$$
C_0 =
 \frac{1}{\sqrt{\pi}}
 \int_{-\infty}^{\infty} d y
  \int_{- 6 }^{-3} dx
  \exp(-y^2)
  \exp(x-k \bar{N}_{init} (x+6)^3 )
   \frac{k^2}{2} y^2 (x+6)^6 2 \bar{N}_{init}
$$
Here we use
$\bar{N}_{eff}(z=-3) = \exp(-3)$
for the number
of droplets  in the unit of the system
because namely this
value corresponds to the characteristic
unperturbed value of droplets and the choice
$$
\bar{N}_{init} = \exp(-6)
$$
which was used in the last formula.

% fail pp1.for iz "special"

The numerical calculations gives
the value of shift $\delta z $
$$
\delta z = \frac{C_0}{\exp(-3) V} \approx
\frac{0.01}{V}
$$

The shift in the total number of droplets
can be easily
calculated since
$$
<N_{tot}> =
 \exp(\delta z ) \bar{N}_{tot}
 $$
 Hence
 $$
 <N_{tot}> \approx \bar{N}_{tot} (1 + \frac{w}{V})
 $$
 $$w \sim 0.01$$

Quite analogously
one can show the smallness of corrections
in a three cycle scheme in the situation of decay.

The result of performed calculations
shows the smallness of
corrections to the
mean value of the total number of
droplets appeared in the processes of
nucleation under the
conditions of decay and
under the smooth behavior of
external conditions.

Correction terms
calculated by the theoretical derivations
in the three cycle scheme
 are so small that they
 can not be confirmed both by
numerical simulations and
experimental researches. So,
generally speaking one has to
say that these stochastic
corrections don't appear
in any practically significant
terms of decompositions.

This result shows that
there is no sense to fulfill the
procedure
of renormalization analogous to that used in
\cite{statiae} in calculation
of dispersion in terms of two
cycle explicit model with a fixed boundary.

Estimating the total result  we have to stress
that the weak feature
of the presented method is the
approximate knowledge of the
back side of the size spectrum
$\exp(-x^3)$
in decay and $\exp(x - (x+3)^3/3^3)$ in the
smooth variation external conditions.
So, these constructions can not be appreciated
as the concretely determined result.

\section{Several first droplets}

The leading
correction term doesn't depend on the volume of
the system. It means that the deviation
in the  mean number of droplets from the
value predicted by TAC doesn't increase with
increase of the volume.
The latter means that namely the first
several droplets are
responsible for initiation of
corrections to the mean value of
droplets. So, now we shall
show how many   droplets are
responsible for
these corrections.

We know that on one hand the Gaussian distribution
can not be applied as
statistics for the first droplets
and on the other hand the
big droplets are extremely
important in kinetics, which
lies in the base of iteration
method. So, it is
reasonable to use the explicit
numerical simulation to see
the role of several first droplets.

We start consideration of the role
of several first droplets with
the case of decay. Figure 4 illustrates
the role of the stochastic appearance
of the first droplet in nucleation
kinetics. Here the relative excess of
mean droplets number is shown.

% file ris4a.mws
\begin{figure}[hgh]

\includegraphics[angle=270,totalheight=8cm]{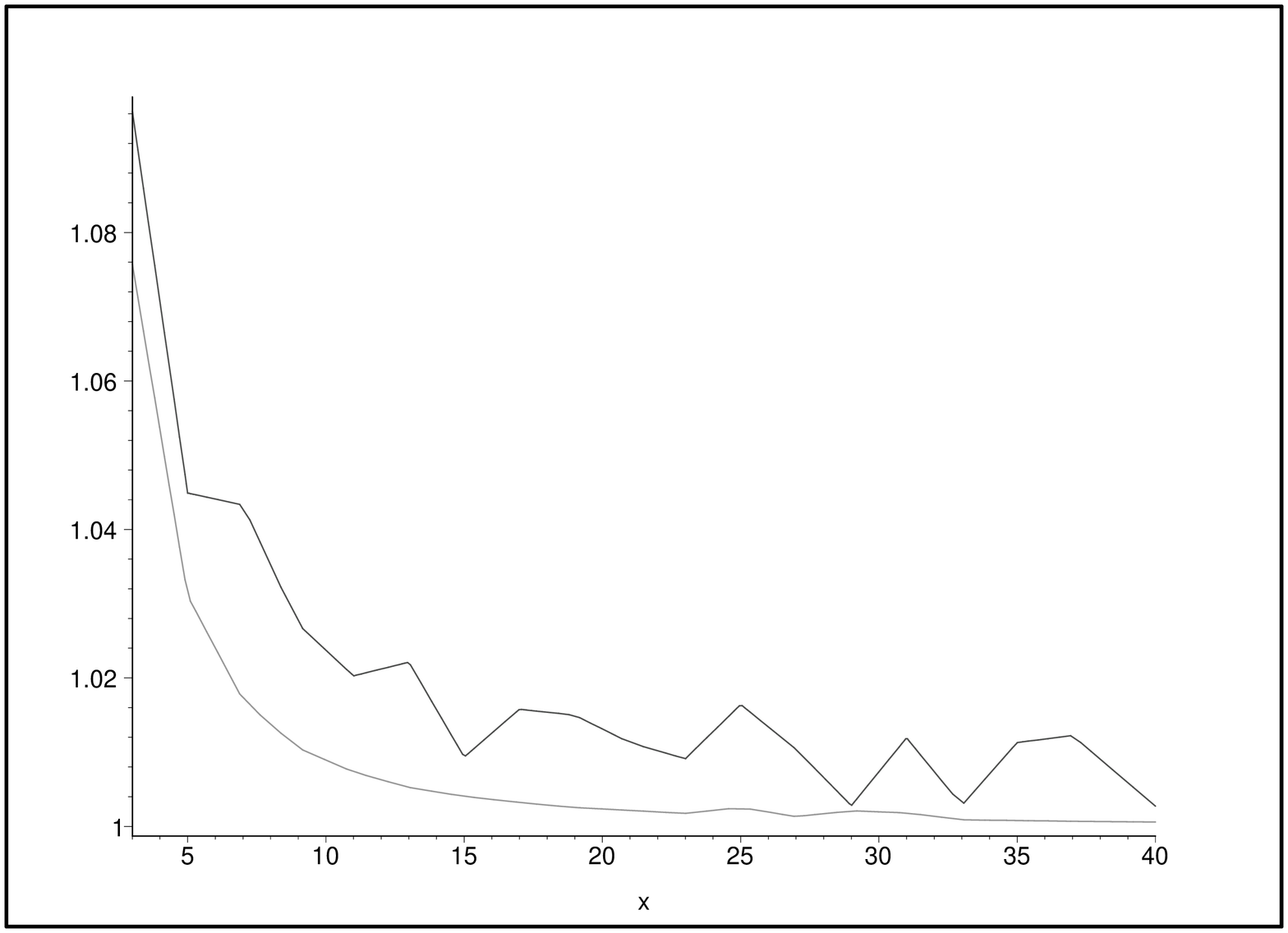}

\begin{caption}
{
Relative excess of the mean droplet number.
Situation of decay}
\end{caption}

\end{figure}

There
are two
curves, both are functions  of the
volume of system $V$. The value of $V$ is
connected with the total number of droplets in TAC
as $ N_{TAC} = 1.28 V$.

The broken line is the result
of numerical simulation for initial
problem, the smooth line is the result
of solution of the following problem:
The first droplet appears
stochastically and later all other
droplets appear with probability
$$
p dx \sim  I  (\frac{dt}{dx}) dx
$$
Here $I$ is the rate of nucleation.
So, except the first droplet the
further appearance occurs according
to  TAC.

One can see the satisfactory coincidence
between the model and the simulation of
initial problem.

One has to note that we are interested
in corrections to the droplets number
when they are essential. We are not
interested in the tails of asymptotics.

The next picture illustrates
the model with two stochastically
appeared droplets. The broken line is
the numerical simulation and the smooth
line is the model with two
stochastically appeared droplets.

% file ris5a.mws
\begin{figure}[hgh]

\includegraphics[angle=270,totalheight=8cm]{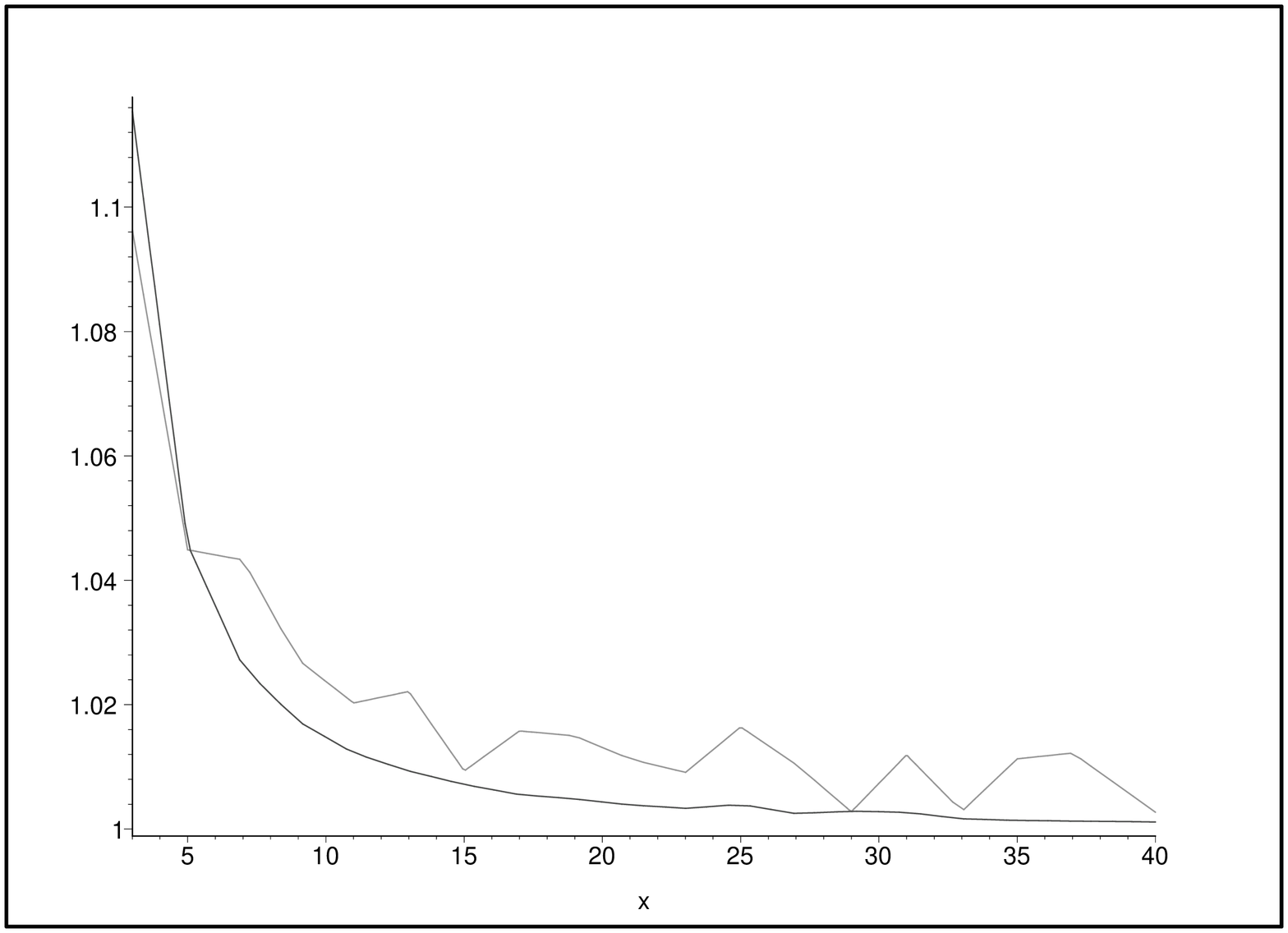}

\begin{caption}
{
Relative excess of the mean droplet number.
Situation of decay. The model with two stochastically
appeared droplets.}
\end{caption}

\end{figure}

Here the coincidence between the
model and simulation is practically
perfect. But  the model with the
first stochastically appeared droplet
is suitable also and due to simplicity has
to be considered as the basic
theoretical model explaining
the corrections to the mean number of droplets.

One can also investigate the model with
discrete regular appearance of
droplets. One can adopt that all
droplets appear when
$$
I_{tot} =
\int_0^t I(t') dt'
$$
attain integer values.
In this model one can take that the
first droplet appears stochastically.
Nothing will be changed.

The results are shown in figure 6.
The axis are the same. One
can see that  the lower broken line
which is the result
of the last model has nothing in
common
with the upper line which is the result
of simulation. So, we has to conclude
that the discrete effects don't
manifest themselves in nucleation
kinetics.

% file ris6a.mws
\begin{figure}[hgh]

\includegraphics[angle=270,totalheight=8cm]{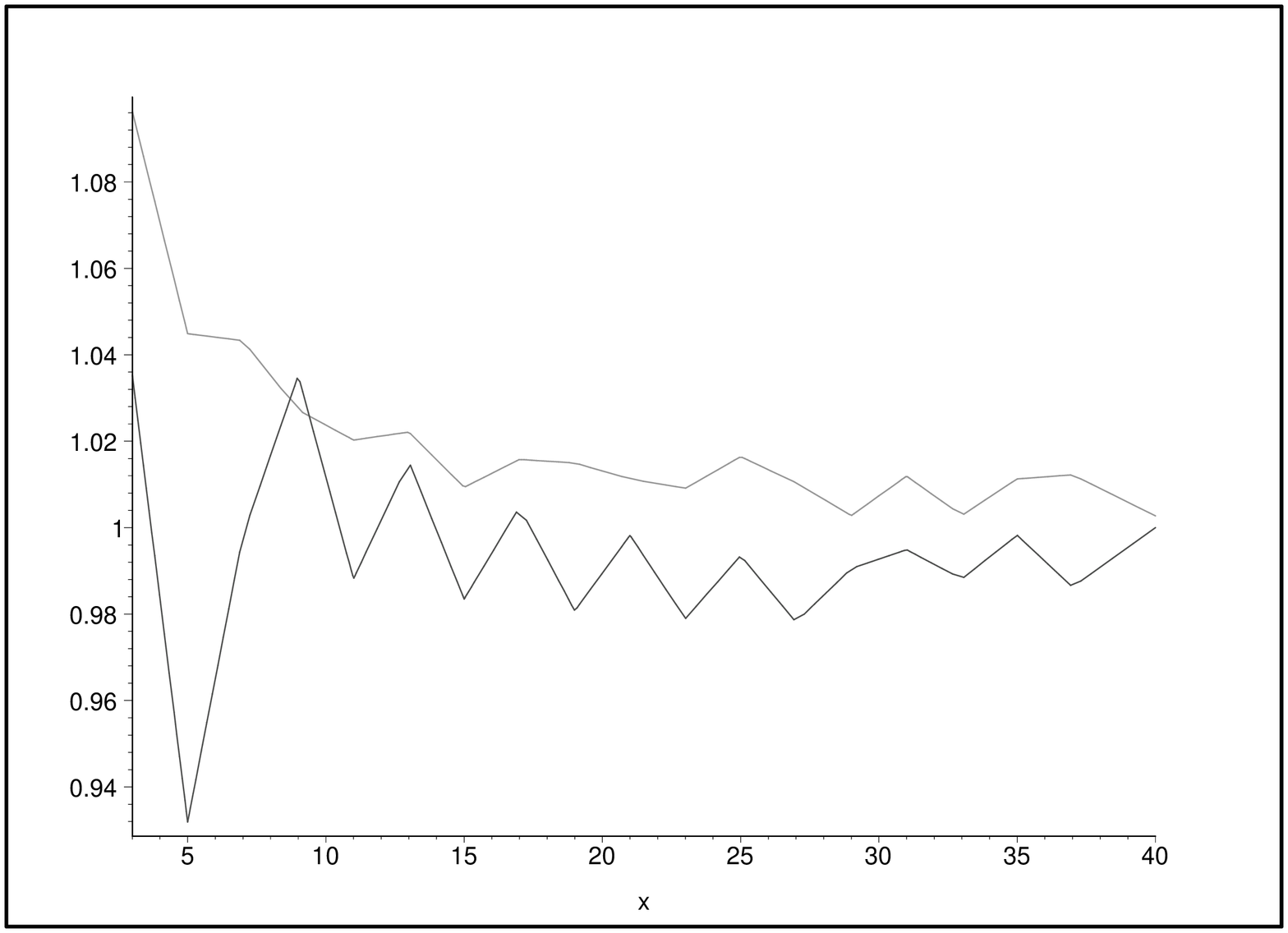}

\begin{caption}
{
Relative excess of the mean droplet number.
Situation of decay. The discrete model and simulation
of initial problem.}
\end{caption}

\end{figure}

One can observe one interesting
feature of kinetics. If in the first
moments of nucleation period the number
of appeared droplets is higher than the
average value then the total number of
droplets will be lower than the average
total value of the droplets number.
This effect will take place at rather
big value of the total number of
droplets. At the small numbers of the
average
total number of droplets the effect
will be the opposite one.

Now we shall turn to investigation of
the nucleation under the smooth
external conditions.

Figure 7  shows results of regular continuous
solutions with several
first droplets born stochastically. There
are three curves drawn in this figure.
The oscillating  curve is numerical
solution,
two smooth curves are approximations with
the only first
droplet born stochastically
and with the first two droplets born
stochastically.
It is clear that there is no big difference
between these  curves.
It means that it is sufficient to
take into account only the
stochastic appearance of the
first droplet.

% fail numdr7.mwz

\begin{figure}[hgh]

\includegraphics[angle=270,totalheight=10cm]{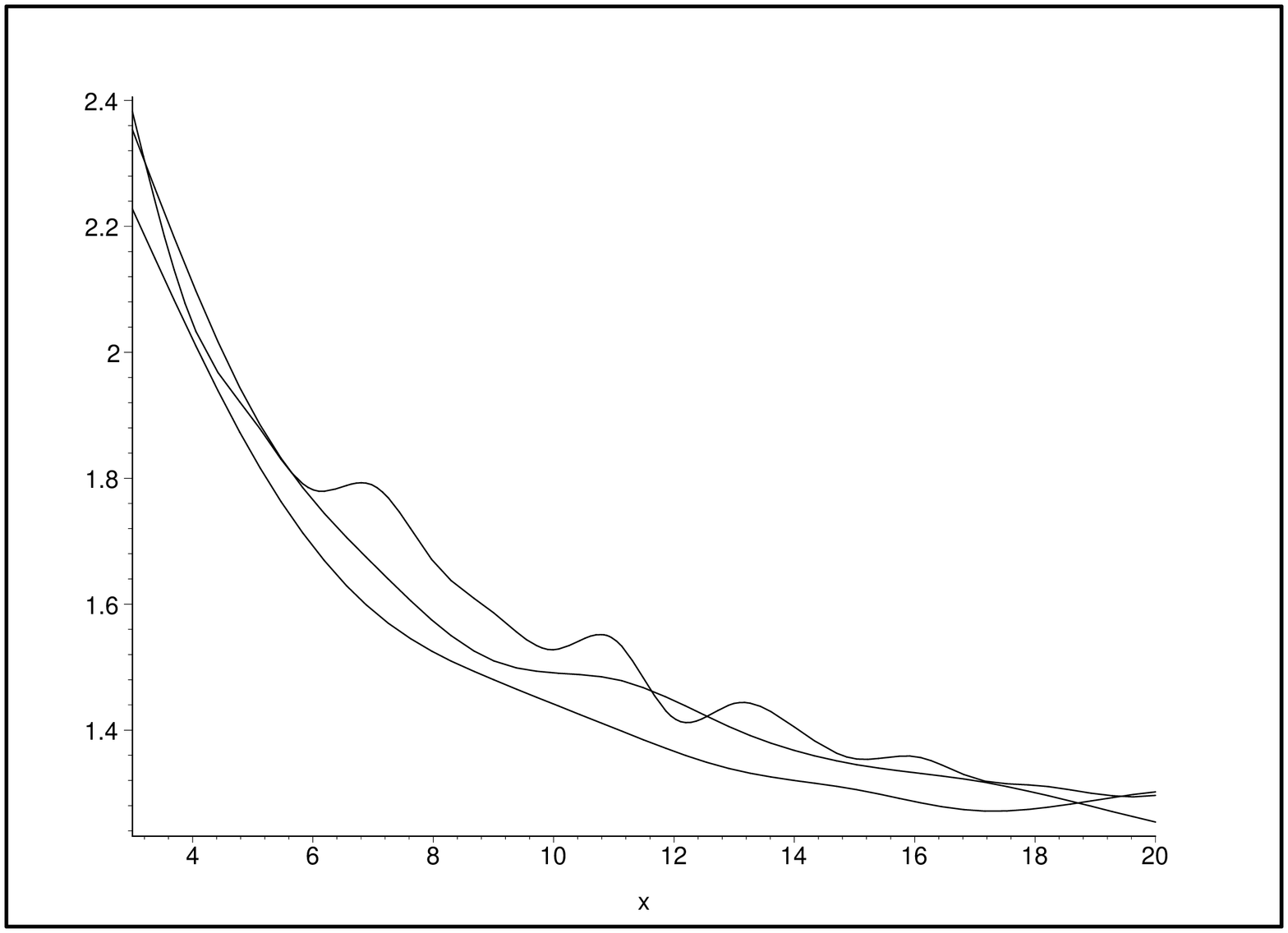}

\begin{caption}
{
Numerical
solution and stochastic approximations at small $V$.
Smooth external conditions.
}
\end{caption}
\end{figure}

We see that the coincidence is rather satisfactory.

Now we shall see how one can incorporate the
discrete
effects in this situations.
We propose the following model.
The first droplet
appears stochastically and later
droplets can
appear only after the elementary fixed
intervals.
Every interval  is
chosen to have  the integral of the rate
of nucleation over
time equal to $V^{-1}$. The
vapor is consumed
by the finite (big) number
of droplets  born in the
mentioned moments of time and growing
regularly.
The result at small $V$ is shown in
Figure 8.

% fail numdr9.mwz

\begin{figure}[hgh]

\includegraphics[angle=270,totalheight=10cm]{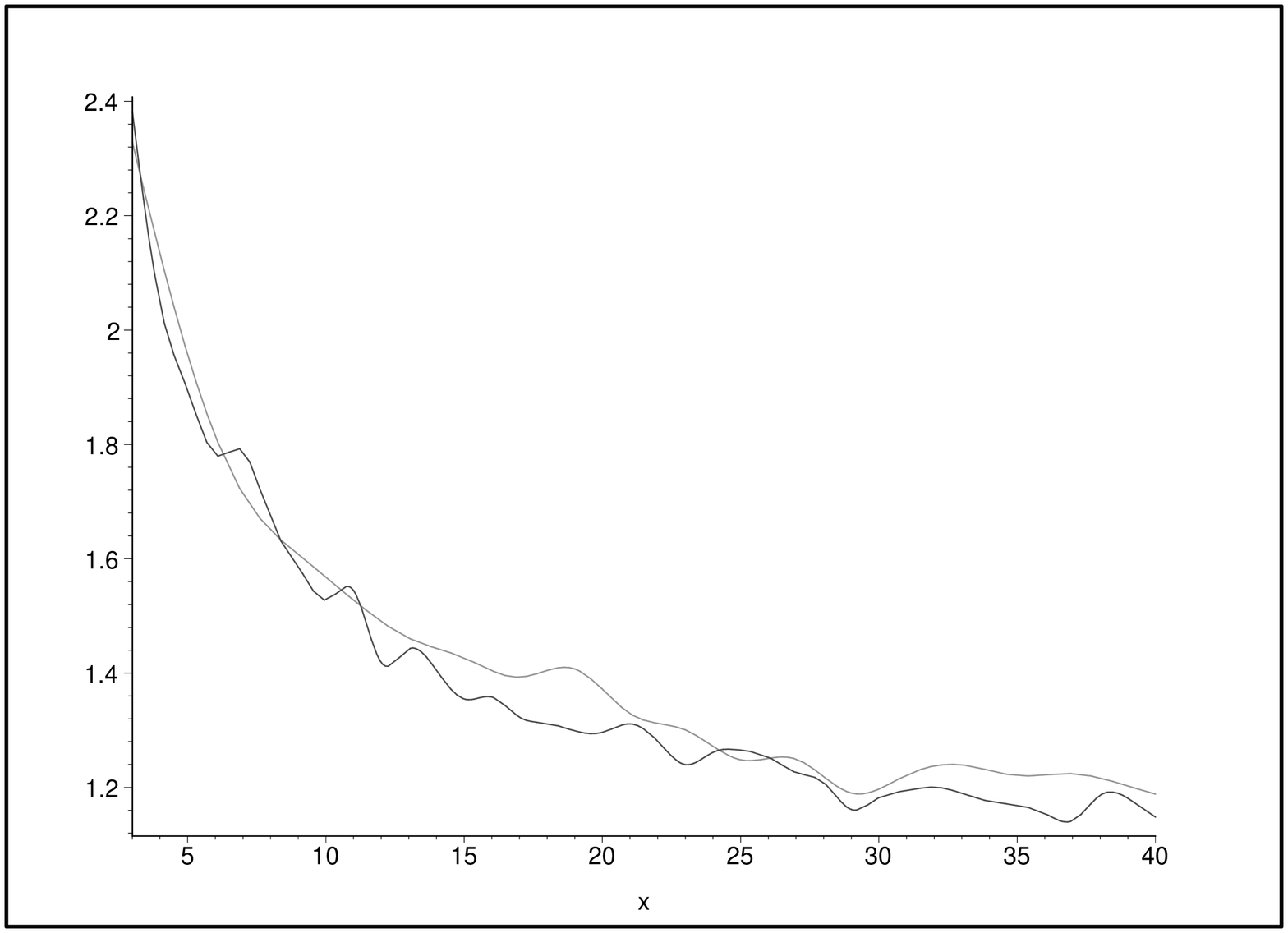}

\begin{caption}
{
Numerical solution and stochastic
discrete approximation at small $V$.
}
\end{caption}
\end{figure}

We see that the
coincidence is satisfactory. May be it is
even better than the
result of the regular continuous model
with a first stochastically
appeared droplet. In any case
we see that the stochastic  appearance of the
first stochastically  droplet
 diminishes the role of
 discrete effects.

One can see that the deviation between discrete and
continuous models is
not big, moreover we see that the role
of discrete
effects is not essential in the deviation of the
average number of droplets.

In this point the nucleation under
conditions of decay differs from the
nucleation under the smooth behavior of
external conditions.

The main result of
performed simulations is that the
stochastic deviation of
mean value of droplets is mainly
caused by the stochastic
appearance of the first droplet.
The stochastic appearance of the
first droplet is very
simple to calculate analytically.
Really, we have to write the Poisson
distribution for the
probability of appearance of the first
droplet
$$
P \sim \exp(-l)
$$
or more concretely  for staying
without appearance of any droplet.
Here $l$ is the number of
possible events. Now we have
to come from $l$ to the time $t$.
This connection is given by
$$
l = \exp(t)
$$
in appropriate renormalization
of time $t$. Certainly this
connection corresponds
to the linearization of ideal
supersaturation
(see \cite{PhysRevE94})
and the ideal rate of nucleation will be
like $\exp(x)$.
Until the
appearance of the first
droplet the supersaturation is
certainly the ideal one.
Then it is easy to get the
differential distribution $p$
over $t$ as
$$
p dt = P dl
$$
Then
$$
p = \exp(t) \exp(-l) = \exp(t - \exp(t))
$$

It is remarkable that the
last distribution is the same as
the universal distribution of droplets in  TAC
established in  \cite{TMF}.

One can write the Poisson distribution
for the appearance of
the first droplet
$p_1 \sim l^1 \exp(-l)$, for appearance of the
first two droplets
$p_2 \sim l^2 \exp(-l)/2$, for
appearance of $n$
droplets $p_n \sim l^n \exp(-l) / n!$, etc.
When $n \ll N_{tot}$
one can use $l=\exp(t)$ to
recalculate $p(t)$ on the base
of $P(l)$. This restriction isn't essential because
at least
$N_{eff} \ll N_{tot}$ and one needs $n < N_{eff}$.

The effects of
discrete model can be also described
analytically. It is
simply necessary to substitute in TAC
the integral by the sum.
One can act in two ways.

The first possibility is to
take explicitly
into account
several first droplets (let it be $K$). Then
the number of droplets in a liquid phase will be
$$
g = \sum_i^K (z-x_i)^3 +
f_* \int_{z_c}^{z} (z-x)^3 \exp(x-g(x)) dz
$$
where $f_*$ is the "amplitude of spectrum"
(see \cite{Novosib}) and $z_i
\equiv x_i $ are determined by
$$
f_* \int_{-\infty}^{z_i}
\exp(x)dx \equiv \exp(z_i) = (i+0.5)
$$
and
$$
f_* \int_{-\infty}^{z_c}
\exp(x)dx \equiv \exp(z_c) = (i+1)
$$

Then the methods of solution
are quite analogous to
\cite{Novosib}.

Another possibility is to use
the Euler-McLorrain decomposition
for
$$
\sum_{i=1}^K z_i - \int_{-\infty}^{z_c} \exp(x) dx
$$
This approach leads to Bernoulli numbers  and
will be published separately.

Also one can use use discrete approximation for all
droplets and replace
it by the integral with the help of
the
global
Euler-McLorrain decomposition.

\section{Concluding remarks}

Generally speaking the
 most important result of the
 given consideration is the
zero correction in the main
term of the shift of the mean
droplets number. The consequence
is the conclusion that
only several few droplets are
responsible for corrections
in the mean number of droplets.
The fact that only several
first droplets form correction in the
total number of
droplets is rather important for
applicability of proposed
method to calculate corrections.
The use of monodisperce
approximation with a fixed boundary is
possible only in the
case when the first correction term is
the zero one and the
first nonzero term  corresponds to the
finite (independent
on $V$) absolute shift in the number of
droplets. Only then
the shift is initiated by
several first droplets and there
is no difference whether we take
into account the rest
droplets in the monodisperce peak or not.
One can note
that the difference between the fixed
boundary and the
floating boundary is reduced only
into account of the rest
droplets. So, there is no difference
what type of boundary
is used (this isn't true for  other
characteristics like
dispersion).

Nevertheless we shall give the corresponding
derivation in
frames of floating boundary.
This is done to show the role of
non-gaussian effects (the
distribution isn't the gaussian one).
Having written the expression
for $<N_{tot}>$
$$
<N_{tot}>
=
\frac{\int_{-\infty}^{\infty} dy
\int_{-3+ y}^{\infty} dx
\exp(x-\frac{1}{27} (x - (3+y))^3 ) dx P(y)}
{
\int_{-3+ y}^{\infty} dx
\exp(x-\frac{1}{27} (x - (3+y))^3 ) dx }
$$
where $y$ is the shift and $P(y)$ is the partial
distribution over coordinate $y$,
 one can get corrections for $<N_{tot}>$.

For the partial distribution $p(y)$ one can write
$$
p(y) dy  = P(N) dN
$$
where $P(N)$ is the
partial distribution over possible
droplets $N$.
Having
written
$$
dN /dy = d\exp(N) / dy = \exp(N)
$$
one can get
$P(N)$.
For $P(N)$ one can write the ordinary gaussian
distribution
$$
P(N) \sim
\exp(-\frac{(N-<N>)^2}{2<N>} )
$$
After transformations we see that
$$
P(N) =
\exp(-\frac{(\exp(-3-y)-\exp(-3))^2}{2\exp(-3)} )
$$
or
$$
P(N) =\exp(-\frac{\exp(-6)
(\exp(-y)-1)^2}{2\exp(-3)} )
$$
Having fulfill decompositions we get
$$
P(N) =\exp(-\frac{\exp(-6)
(1-y+y^2/2+ ... -1)^2}{2\exp(-3)} )
$$
and with restriction of first terms
$$
P(N) =\exp(-\frac{\exp(-6)
(y-y^2/2)^2}{2\exp(-3)} )
$$
or finally
$$
P(N) =\exp(-\frac{\exp(-6)
y^2}{2\exp(-3)} )
\exp(-\frac{\exp(-6)
y^3}{2\exp(-3)} )
$$
Having extracted Gaussian distribution we get
$$
P(N) =\exp(-\frac{\exp(-6)
y^2}{2\exp(-3)} )
(1-\frac{\exp(-6)
y^3}{2\exp(-3)} + ...)
$$
So, there appear the non-gaussian corrections.
Namely  these corrections are
the reason of appearance of
corrections in the total number of droplets.
They can be easily
calculated by the manner described in
calculations in the model with a fixed boundary.

Here we shall stop our
calculations and put a question what
distribution has
to be a gaussian one: the distribution
$P(N)$ or the distribution $P(y)$?
Certainly there
is no clear answer on this question.
Moreover the
results of \cite{statiae} shows that there
is a real difference
 when  gaussian distribution instead of
the
Poisson distribution is used.
Here it is clear that we
have to use the Poisson distribution.
But then  to fulfill the integration
one has to use the steepens
descent method which is equivalent to the use of
the gaussian
distribution with corresponding corrections.

So we came to a paradox and
it can not be resolved without
taking into account that several first
droplets are the
reason of the shift of the mean number of droplets.
Fortunately, there is
no need to continue this procedure
and one can take into
account the influence of several first
droplets
explicitly by the procedure described above.

In investigation of the
shift to the droplets number one has
to take into account
that the asymptotic we need is the
"intermediate asymptotic".
There is no necessity to know
for example that instead
of $10000$ there will be $10005$
droplets. We need the
shift where it is at least few
percent.
So, we need asymptotics at the intermediate mean
number of droplets.
Namely this case was investigated and
it was shown that already account of two or
three first droplets
is
sufficient for the true  shift of the
droplets number.

We have to note that there is another reason of
applicability of
monodisperce approximation with a fixed
boundary. This reason lies in
construction of monodisperce
approximation and it is
different for decay and for smooth
variation of
external conditions. For the situation with
the smooth variation of
external conditions one can note
that the
amplitude $\exp(-3)$ corresponding to the moment
of
formation of monodisperce peak is very small. So, the
value $y \exp(-3)$ will be
small. Then we can neglect
$y$ in the
lower boundary of integration and come to the
approximation with a
fixed boundary instead of approximation
with a floating boundary.
So, the smallness of $\exp(-3)$ is the
reason why one can
use approximation with a fixed boundary.

In the situation of
decay there is no such smallness of
amplitude. But one can recall that
the shift of monodisperce
approximation (i.e. the
position of peak formation) was chosen
in such a way that "the length" of
peak corresponds to the
extremum of droplets
number (see \cite{Decaydispersion}).
So, the derivative of
the total number of
droplets over the
length of
spectrum is zero and there is no difference
whether to use the
fixed boundary or to use the floating
boundary.
Here appears the physical reason of the choice
of monodisperce
approximation in a way prescribed in
\cite{Decaydispersion}.
Certainly neither the approximation
of fixed boundary nor
approximation of floating boundary
reflect the right
physical evolution but such a choice of
monodisperce approximation
allows to ignore this problem.

\end{document}